\documentclass[aps,prl,twocolumn,floatfix,showpacs]{revtex4}
\usepackage{graphicx}
\begin{document}
\title{Plume motion and large-scale circulation  in a cylindrical Rayleigh-B\'enard cell} 
\author{Denis Funfschilling}
\author{Guenter Ahlers}
\affiliation{Department of Physics and iQUEST,\\ University of  California, Santa Barbara, CA  93106}
\date{ \today}

\begin{abstract}
We used the time correlation of shadowgraph images to determine the angle $\Theta$ of the horizontal component of the plume velocity above (below) the center of the bottom (top) plate of a cylindrical Rayleigh-B\'enard cell of aspect ratio $\Gamma \equiv D/L = 1$ ($D$ is the diameter and  $L \simeq 87$ mm the height) in the Rayleigh-number range $7\times 10^7 \leq R \leq 3\times 10^{9}$ for a Prandtl number $\sigma = 6$. We expect that $\Theta$ gives the direction of the large-scale circulation. It oscillates time-periodically. Near the top and bottom plates $\Theta(t)$ has the same frequency but is anti-correlated. 
\end{abstract}

\maketitle
Turbulent Rayleigh-B\'enard convection (RBC) is an important  process that occurs in the oceans, the atmosphere, the outer layer of the sun, the Earth's mantle, and in many industrial processes.\cite{SD01} Despite the seeming simplicity of the idealized laboratory experiment, i.e. a fluid between horizontal parallel plates heated from below and cooled from above, our understanding of the physical mechanism of turbulent RBC  remains incomplete.\cite{Si94,Ka01,AGL02} Much of the heat transport is mediated through the emission of hot (cold) plumes of fluid from a thin boundary layer adjacent to the bottom (top) plate, and these plumes are carried by a large-scale circulation \cite{KH81,HCL87,CGHKLTZZ89,CCS97,QT01,NSSD01, QT02} known as the ``wind of turbulence". For systems with aspect ratio $\Gamma \equiv D/L = {\cal O}(1)$ ($D$ is the diameter and  $L$ the height of a cylindrical cell) this wind, when time averaged, takes the form of a single convection roll filling the entire cell. The plume interaction with this circulation is a central component of the turbulent RBC problem, and yet only little is known quantitatively about plume motion and the wind. To a large extent we expect that at least the {\it horizontal} motion of the plumes is slaved to the wind velocity, since the only independent force acting on the plumes is the buoyancy force in the {\it vertical} direction. Here we present a quantitative study of  plume motion and thus, by inference, of the wind direction above (below) the center of the bottom (top) plate. The measured horizontal direction oscillates time periodically, and the oscillations persist with a unique frequency over hundreds of cycles. The oscillations at the top and bottom have the same frequency but a phase which is displaced by half a cycle. We conclude that the wind is a significantly more complicated dynamical system than a simple Rayleigh-B\'enard convection cell.

Numerous measurements of the speed of the fluid and/or of the temperature at distinct points (i.e. of scalar quantities) were made before and revealed a time-periodic component. \cite{HCL87,CGHKLTZZ89,CCS97,QT01,NSSD01,QT02} In some instances these results were interpreted as indicative of a time-periodic plume emission from the plates. Where direct comparison is possible, we find that the measured frequencies agree quantitatively with our determination of the oscillation frequency of the horizontal wind direction. It has been shown \cite{QT01} that the corresponding oscillation period is commensurate with the period of circulation of the wind. It had been suggested that periodic plume emission from the top and bottom plates, synchronous with the wind circulation period, provided a mechanism for driving the wind. Our results offer a different explanation of the periodicity of the system, and thus there is no longer a direct reason to assume that {\it periodic} plume emission is a central ingredient of the driving mechanism.

Rayleigh-B\'enard convection is characterized by the Rayleigh number 
$R = \alpha g \Delta T L^3/\kappa \nu$
and the Prandtl number 
$\sigma = \nu/\kappa$.
Here $\alpha$ is the isobaric thermal expansion coefficient, $\kappa$ the thermal diffusivity,  $\nu$ the kinematic viscosity, $g$ the acceleration of gravity, and $\Delta T$ the applied temperature difference. We used methanol with  $\sigma = 6.0$. For $0.5 \alt \Delta T \alt 20^\circ$C at a mean temperature of 40$^\circ$C they covered the interval $7\times 10^7 \leq R \leq 3\times 10^{9}$, in a range sometimes called the "hard turbulence" regime \cite{HCL87}, but below the Kraichnan regime \cite{Kr62} where the boundary layers themselves become turbulent and break up. 

The apparatus was described before.  \cite{DAC95,XBA00,AX01} The cell had a sapphire top plate of thickness 0.32 cm. The bottom plate, of aluminum 0.65 cm thick, had a top surface which was polished to a mirror finish and was heated from below by a metal-film heater uniformly distributed over the cross section of the cell. The wall was made of high-density polyethylene. The internal cell diameter and height were $D \simeq 87.4$ and $L \simeq 86.6$ mm respectively. 

\begin{figure}
 \includegraphics[width=6cm]{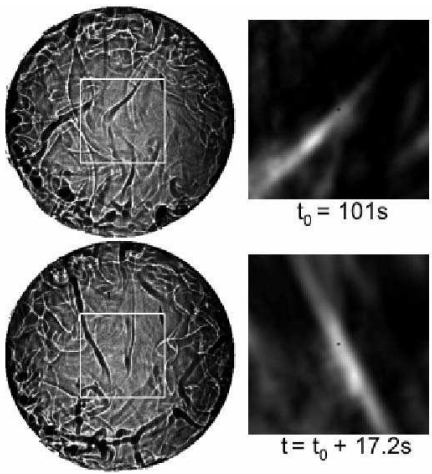}
\caption{Shadowgraph images of plumes for R = $5.6\times10^8$. The elongated black stripes are the hot plumes just above the bottom plate. The white squares in the images, corresponding to $31\times31$mm$^{2}$, indicate the part used to calculate the temporal correlation between two successive images separated in time by 1.22s.  The auto-correlation functions are shown on the right. A black dot at their center marks their origin.}
\label{fig:correlation}
\end{figure}

Above  the cell was a shadowgraph optical system \cite{DBMTHCA96}. Shadowgraph images represent a vertical average of the horizontal Laplacian of the refractive index (and thus temperature) field in the cell. In the central part of the cell, between the thermal boundary layers, the temperature undergoes vigorous fluctuations but on average is nearly equal to the mean temperature of the system. This part of the cell does not contribute significantly to the shadowgraph image. The {\it laminar} boundary layers themselves also do not contribute because they have no significant horizontal temperature variation. Thus the major contribution to the images comes from the plumes which emerge from the bottom or top boundary layers. These consist of localized relatively warm or cold fluid respectively, and thus appear as dark or bright areas. In order to reduce contributions from optical imperfections of the system, the images are divided by a reference image taken with $\Delta T = 0$ and then rescaled. Two typical divided images  are shown in the left column of Fig.~\ref{fig:correlation}. With the particular setting of the optics, the warm dark plumes at the bottom are more apparent than the cold bright ones at the top, but both are present. In an attempt to break the rotational invariance of the system and to give the large-scale flow a preferred direction, our measurements initially were made with the entire system tilted by an angle of 2$^\circ$. However, a subsequent sequence of measurements with the system horizontal within 0.1$^\circ$ gave the same results. Using camera exposure times of 30 or 60 ms, we took sequences consisting of 8192 or 16386 images at time intervals of 200 to 500  ms, corresponding to an acquisition time of  30 to 150 min or 50 to 350 periods of  the oscillations, for a given $R$.

Plumes were visualized from the side many times, either by the shadowgraph method \cite{ZML90,MZL93,CCL96,SQTX03} or by using thermochromic liquid crystals.  \cite{ZML90,QT01b}. Those observations suggest that plumes have the shape of a mushroom with a cap, and with a stem connecting the cap to the boundary layer. We find that visualization from the top reveals new features. \cite{ZML90} In Fig.~\ref{fig:correlation} one sees that the plumes are extended horizontally, consistent with rising or sinking sheets of fluid. We only very rarely observed a typical mushroom cap at the lateral termination of a plume. The plumes in our experiment have a horizontal extent  that varies from a few mm to a few cm, and a thickness of around 1 to 2 mm for R = $2\times10^8$. This plume thickness is about the same size as the thermal boundary-layer thickness, as expected. \cite{CG73,CL99}  Comparison of successive images, separated in time by a second or so, shows that the plumes are moving, somewhat erratically, but on average in the direction of their long axis. The alignment of convection rolls by a prevailing wind or by shear flow is a well known phenomenon.\cite{Ku71,Ke77}.
 
Visual observation revealed that the direction of the plume movement oscillated in time. We calculated time auto-correlation functions between successive images. First, a threshold was applied and a binary image was obtained, rendering the dark plumes in black. Then, on a central part of the image of size $31\times31$mm$^{2}$, illustrated by the squares in the images of  Fig.~\ref{fig:correlation}, the correlation between successive images was computed. Two examples of these correlation functions are shown in the right column of Fig.~\ref{fig:correlation}. The vector from the origin to the highest peak gave the average displacement of the plumes which occurred during the time interval between the 2 images, and thus the plume velocity. Here we focus on the angle $\Theta$ of this vector relative to an axis which was chosen so that the time average of $\Theta$ vanished.

\begin{figure}
 \includegraphics[width=6cm]{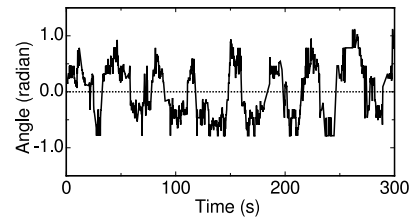}
\caption{Temporal evolution of the angle $\Theta$ of the horizontal velocity component of the plumes close to the bottom plate. As for Fig.~\ref{fig:correlation}, R = $5.6\times10^8$.}
\label{fig:angle}
\end{figure}

\begin{figure}
 \includegraphics[width=6cm]{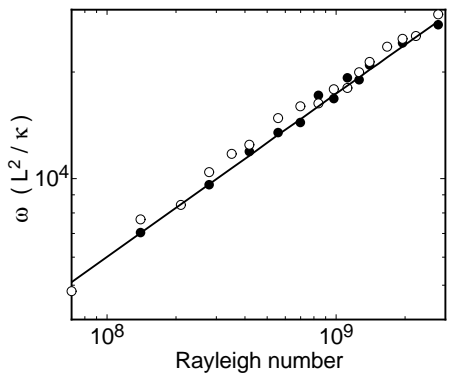}
\caption{The oscillation frequency $\omega$, in units of the vertical thermal diffusion time $L^2/\kappa$, of the angle $\Theta$ of the plume motion just above the bottom plate for the slightly tilted (open circles) and the untilted (solid circles) apparatus. The straight line correspond to a fit of a powerlaw to the data of Ref. \protect \cite{QT01}. }
\label{fig:frequency}
\end{figure}

Results for $\Theta$ are shown in Fig.~\ref{fig:angle}. As already noted by visual observation, $\Theta$ oscillated in time with a well defined frequency $\omega$. The oscillations covered an angular range of more than $90^\circ$. Results for $\omega$, scaled by the vertical thermal diffusion time $L^2/\kappa = 7.88\times10^4$ s,  are shown in Fig.~\ref{fig:frequency} as a function of $R$  for both the slightly tilted (open circles) and the untilted (solid circles) apparatus. The two sets agree quite well with each other, and can be described by $\omega = \omega_0 R ^ {-\beta}$, with $\omega_0 = 1.18 (1.26)$ and $\beta = 0.465 \pm 0.011 (0.460 \pm 0.012)$ for the tilted (untilted) run. 

The existence of time-periodic behavior in our range of $R$ was observed previously, by measurements either of the temperature or of a component of the velocity (i.e. of scalar quantities) at
 isolated points in the convection cell, in cylindrical \cite{HCL87,CGHKLTZZ89,CCS97,QT01,NSSD01,QT02} as well as rectangular \cite{CCL96} systems. These measurements could determine the 
frequency, but could not identify the nature of the oscillating mode. The measured frequencies depended on $R$ (as shown in Fig.~\ref{fig:frequency}) and
$\sigma$. 
For the same $\sigma$, our results agree quantitatively with those obtained by the scalar 
measurements,\cite{QT01} as illustrated by the solid line in Fig.~\ref{fig:frequency} which represents a fit $\omega = 1.26R^{0.46}$ to data with  $\sigma = 5.4$.\cite{QT01}
Thus the scalar and our vector measurements capture the same 
physical phenomena. In some cases this frequency was attributed to periodic plume 
emission \cite{SWL89,V95,QT02,QSTX03}. Our results indicate that its origin is the horizontal oscillation of the large-scale flow. 

It had been determined by others that, consistent with a model given in Ref. \cite{V95},  the frequency is proportional to the Reynolds 
number $R_e$ of the large-scale circulation.\cite{CCS97,QT01,SWL89,QYT00,LSZX02} The prediction for $R_e$ by Grossmann and Lohse \cite{GL01}, for our $\sigma$ and over our range of $R$,
gives $\beta$ as an effective exponent with a value of 0.446, slightly lower than but reasonably consistent with our measurements and with those of Ref. \cite{QT01} .

\begin{figure}
 \includegraphics[width=6cm]{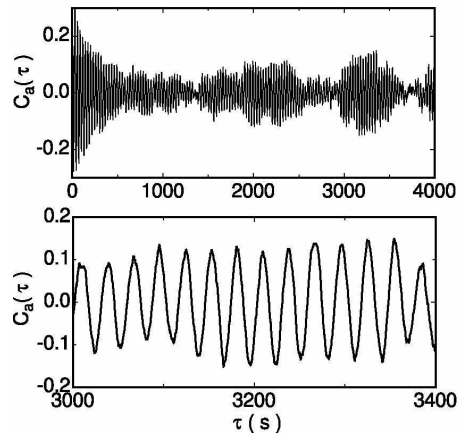}
\caption{Auto-correlation of the angle $\Theta$ for $R = 8.37\times 10^8$. The lower panel shows a small time interval of the upper one on an expanded scale.}
\label{fig:autocorrelation}
\end{figure}

We obtained the auto-correlation function 
\begin{equation}
C_a(\tau) = \langle \delta\Theta( t )\delta\Theta( t + \tau) \rangle / \langle \delta\Theta(t)^2\rangle
\end{equation}
from image sequences covering typically 8000s.
Here $\delta\Theta( t ) = \Theta( t ) - \langle \Theta(t) \rangle$ (angular brackets indicate a time average).
An example in Fig.~\ref{fig:autocorrelation}  shows some decay for the first 500s, but then continues to oscillate indefinitely with an amplitude which varies irregularly. This indicates that the flow maintains its characteristic frequency indefinitely with high accuracy, but does not rule out a slow phase shift with time.

\begin{figure}
 \includegraphics[width=6cm]{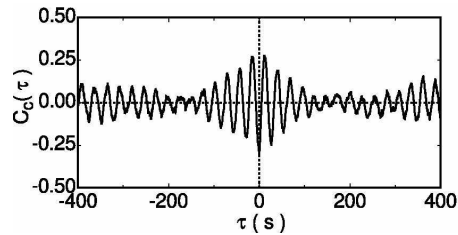}
\caption{Cross correlation between the direction of the hot and cold plumes for $R = 9.77\times 10^8 $.}
\label{fig:phase}
\end{figure}

The shadowgraph apparatus had a depth of field large enough to visualized the hot plumes near the bottom (which appear dark), and the cold plumes near the top (which appear bright). By choosing different threshold values in the image processing, we were able to separate the hot and cold plumes, and determine  the velocity of both of them from the same images. Both velocities have similar oscillation amplitudes and the same $\omega$. An example of the cross-correlation function of the hot- and cold-plume angles $\Theta_{h}$ and $\Theta_{c}$ 
\begin{equation}
C_c(\tau) = \frac {\langle \delta\Theta_{h}( t )\delta\Theta_{c}( t + \tau) \rangle }  {\sqrt{\langle \delta\Theta_{h}^2(t)\rangle \langle\delta\Theta_{c}^2(t)\rangle}}\ .
\end{equation}
is given in Fig.~\ref{fig:phase} and shows that the directions of motion of the hot and cold plumes are anti-correlated, i.e. out of phase by $\pi$. This was the case for all $R$.

\begin{figure}
 \includegraphics[width=5.0cm]{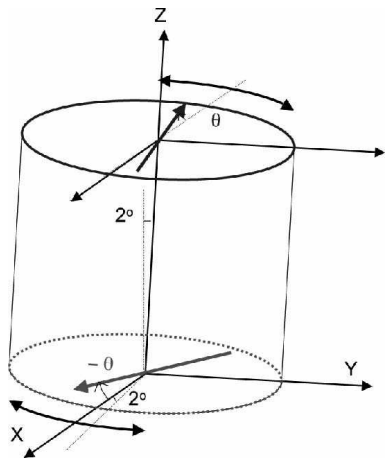}
\caption{Schematic summary of a snapshot of the plume movement. The small angle at the bottom labeled $2^\circ$ indicates the tilt of the apparatus relative to gravity. The curved double-headed nearly horizontal lines near the top and bottom indicate the amplitude of oscillation of the angle $\Theta$. The arrows through the top and bottom plane are an example of the direction of flow at an angle $\Theta$ at the top and $-\Theta$ at the bottom at a given time. Although not directly evident from our measurements, we assume that the instantaneous direction of flow near the side wall is given by connecting the head of the top (bottom) arrow with the tail of the bottom (top) arrow.}
\label{fig:shemat}
\end{figure}

Let us summarize our understanding of the large-scale flow. One of the interesting features of Rayleigh-B\'enard turbulence began to emerge when it was found that the flow, although vigorously fluctuating, highly irregular, and seemingly random, contains a well defined frequency.\cite{HCL87} This frequency is Rayleigh-number dependent, and its period is proportional to a turnover time of a large-scale circulation.  The Reynolds numbers of the large-scale flow implied by the frequency measurements agree well with the prediction of Grossmann and Lohse.\cite{GL01} 
The existence of the oscillations of the flow direction shows that the flow is much more structured than initially assumed. Due to contributions of a number of investigators,\cite{HCL87,CGHKLTZZ89,CCS97,QT01,NSSD01, QT02,CCL96,QYT00,LSZX02} a lot of progress was made in our understanding of the large-scale circulation, its periodicity,  the boundary layers, plume dynamics, and the plume interaction with the large-scale wind. A schematic drawing of the velocity field and the plume distribution had evolved \cite{Ka01} and was named the "flywheel" by some. It showed the fluid circulating as a loop in a well defined plane with a vertical axis, stationary in time, and the plumes rising and falling on opposite sides of the loop. Alhough this picture was appealing, it was called into question already by measurements of the characteristic frequency and of the amplitude of the oscillations at various places in a cell.\cite{QSTX03}

Our work focused on the plumes viewed from the top. Our first result is that  near the bottom and the top plate, plumes when viewed from above look like sheets. They move across the field of view, and are elongated in the direction of their movement. Our second result is that the frequency measured in the temperature or flow field corresponds to a periodic oscillation in the horizontal plane of  the direction of the plume movement, in contradiction to the simple flywheel model. Our third result is that the direction of the hot and cold plumes near the top and bottom plates are out of phase by $\pi$. A simple sketch summarizing the new features of the flow is given in Fig~\ref{fig:shemat} and shows that the schematic picture of a simple loop (flywheel), partly driving  and in turn partly driven by plumes, is an over-simplification. It turns out that the large-scale flow has a more complicated three-dimensional
 spatial structure, and that it may be regarded as an interesting dynamical system with a characteristic frequency of a three-dimensional mode.

This work was supported by the US Department of Energy through Grant  DE-FG03-87ER13738.

\end{document}